\documentclass[runningheads]{llncs}

\usepackage{graphicx}
\usepackage{url}
\usepackage{amsmath}
\usepackage[hidelinks]{hyperref}
\usepackage{cleveref}
\usepackage{physics}
\usepackage{braket}

\usepackage{xcolor}

\begin{document}
\title{The Fine Print of Solving Markov Chains with Analog Quantum Computing}
\titlerunning{Solving Markov Chains with Analog Quantum Computing: The Fine Print}
\author{Ward van der Schoot \and Niels M. P. Neumann}
\institute{Department of Applied Cryptography and Quantum Applications, TNO, Anna van Buerenplein 1, 2595 DA Den Haag, The Netherlands}

\maketitle
\begin{abstract}
     With a growing interest in quantum computing, the number of proposed quantum algorithms grows as well. 
     The practical applicability of these algorithms differs:
     Some can be applied out-of-the-box, while others require black box oracles, which can not always be easily implemented. 
     One of the first works to explicitly discuss these practical applicability aspects is by Aaronson discussing the \textit{fine print} of the HHL quantum algorithm that solves linear systems of equations. 
     We extend this line of research by providing a similar fine print for the first analog quantum algorithm that computes the stationary distribution of Markov chains. 
     We conclude that more focus should be put on this practical applicability of quantum algorithms, either through a separate line of research, or through more attention when introducing the algorithm. 
\end{abstract}

\section{Introduction}
More than thirty years ago, the first quantum algorithms were proposed, including the algorithms by Deutsch-Jozsa~\cite{DJ:1992}, Shor~\cite{Shor_1994} and Grover~\cite{Grover:1996}. 
Since then, an enormous amount of quantum algorithms have been proposed, with varying improvements over conventional methods. 
An interesting algorithm is given by Harrow, Hassidim and Lloyd (HHL), whom proposed a quantum algorithm for solving linear systems of equations~\cite{HHL2009}. 
Interestingly, they claimed an exponential speedup over classical methods in their original publication. 

However, the HHL algorithm produces the solution of the linear system as a quantum state. 
To obtain the answer to use in further conventional processing requires the algorithm to be run multiple times.
Because of this, the exponential speedup vanishes in the general case.
The advantage in practice depends on the specific use-case and the implementation. 
This and other nuances when implementing the HHL algorithm are discussed by Aaronson in the fine print~\cite{Aaronson2015}.

Later, Ref.~\cite{Neumann:2019} gave similar fine print for a clustering algorithm using quantum persistent homology, showing that the claimed advantages vanish in practice.
Other algorithms often suffer from similar nuances, yet they remain largely undiscussed. 
We therefore continue this line of research by discussing a the direct applicability of a recently proposed quantum algorithm and providing the fine print for it. 

Our focus will be on an analog quantum algorithm that can prepare the stationary distribution of a Markov chain as a quantum state~\cite{Chakraborty_2020}.
Their algorithm claims to achieve a complexity similar to the one obtained by state-of-the-art discrete quantum algorithms~\cite{Orsucci2018fasterquantummixing,LiShang:2023}.
The approach assumes the existence of a Hamiltonian that encodes the Markov chain, which is then used in combination with Von Neumann measurements. 
These two primitives are combined to prepare the stationary distribution of the Markov chain.

The work by Chakraborty, Luh and Roland focuses solely on the theoretical description of the algorithm, including an elaboration on the speedup that the algorithm yields. 
Our work considers the algorithm in more detail and considers the low-level specifications.
It is concluded that the theoretical speedup only holds under the assumption that certain variables are available in the application of the algorithm. 
In general, the complexity to compute these variables is higher than the complexity of the algorithm.
More precisely, this complexity is actually the same as the classical time to prepare the stationary distribution, implying that the claimed speedup does not hold.

It is important to note that the aim of this work is not to discredit the work by Chakraborty et al., but rather to put the algorithm in a realistic perspective.
The analog quantum algorithm is still able to prepare the stationary distribution, yet without the claimed speedup.
In addition, the algorithm contains building blocks that could potentially be used to devise an algorithm to prepare the stationary distribution with a speedup over classical methods.
This fine print, however, forms an important guideline regarding the practical applicability of the algorithm for potential end users. 

This work is structured as follows:
First, the next section gives background information on Markov chains and the algorithm by Chakraborty, Luh and Roland~\cite{Chakraborty_2020}.
Then, section~\ref{sec:fine_print} discusses the fine print of the algorithm, with an elaboration as to why the promised speedup does not actually hold in general.
The work finishes with a discussion and outlook in Section~\ref{sec:discussion}.

\section{Background}
\subsection{Markov chains}
Markov chains, first described by Markov~\cite{Original_Markov} and later extended by Poincaré and Kolmogorov~\cite{Kolmogorov}, form a class of basic mathematical models that can describe a wide variety of different processes, for example in network science~\cite{Network_science}, combinatorial optimisation~\cite{combinatorial_optimisation}, statistical physics~\cite{Statistical_physics} and Monte Carlo methods~\cite{Monte_Carlo}. 
In this work only discrete Markov chains are considered, though continuous ones exist as well. 
Such a Markov chain is a discrete, indefinite process $(X_m)_m$, such that each $X_m$ is equal to a certain probability distribution over all possible states in a certain state space. 
In a Markov chain, the probability distribution of the next state depends only on the current state, i.e.,
\begin{equation*}
    \mathbf{P}(X_{m+1}=x_{m+1}|X_{0}=x_0,\ldots X_m=x_m) = \mathbf{P}(X_{m+1}=x_{m+1}|X_m=x_m)    
\end{equation*}
for all positive integers $m$ and $x_i$ probability distributions over the state space.

In homogeneous Markov chains, the probability is independent of $m$, i.e., $\mathbf{P}(X_{m+1}=x_{m+1}|X_m=x_m)=\mathbf{P}(X_{1}=x_{1}|X_0=x_0)$ for all positive integers $m$. 
These Markov chains can be represented as square matrices: if the state space $X=\{1, 2, \ldots , n\}$ of the Markov chain has size $n$, the Markov chain can be represented as an $n\times n$ matrix $P$ such that the entry $P_{ij}$ is the probability to transition from state $i$ to state $j$. 
These probabilities are called the \textit{transition probabilities} of the Markov chain, and the corresponding matrix is called the \textit{transition matrix}.
Note that such a matrix is always a stochastic matrix, as the entries of each row sum to 1.
Now, if the probability distribution of the current state is equal to $v$, written as a vector, then the probability distribution of the next state can easily be computed as a vector $w$, such that 
\begin{equation*}
    w_i = \sum_{j=1}^nP_{ji}v_j
\end{equation*}
which translates to $w=P^Tv$. 
From this matrix, another representation of a Markov chain can be obtained, namely in the form of a directed, weighted graph, with potential self-loops, in which the states are the nodes, and the edges are given by the transition probabilities.

Ergodic Markov chains take a central role in the field of Markov chain research.
Such ergodic Markov chains are both aperiodic --the greatest common divisor of all cycles in the graph of the Markov chain is $1$-- and positive recurrent --the probability of returning to the original state at any moment is 1 for any initial state, and the expected time of returning is finite.
Interestingly, the probability distribution of ergodic Markov chains converges over time to a unique (stationary) probability distribution $\pi$, independent of the initial probability distribution. 
The stationary distribution remains static under the Markov chain, such that $\pi=P^T\pi$.

The \textit{time-reverse} of a Markov chain $(X_m)_m$ is the Markov chain obtained by running the original Markov chain in reverse, i.e., $(X_{T-m})_m$ for some fixed $T$. 
Ergodic Markov chains remain ergodic when reversed, with the same stationary distribution. 
The fact that a Markov chain is time-reversible Markov chains $P$ with stationary distribution $\pi$ is equivalent to the \textit{detailed-balance equations}:
\begin{equation*}
    \pi_iP_{ij}=\pi_jP_{ji}    
\end{equation*}
for all states $i$ and $j$.
This work only considers homogeneous, reversible, ergodic Markov chains.

Many applications in Markov chain theory are interested in either preparing the stationary distribution of a Markov chain, or sampling from it.
Traditionally, the stationary distribution of a Markov chain can be prepared by \textit{mixing} it: repeatedly applying the Markov chain to some initial distribution.
The number of required applications to get $\varepsilon$-close to the stationary distribution is called the \textit{mixing time}.
Recently, with the advent of quantum computing, quantum methods to prepare the stationary distribution have been studied.
Because of the probabilistic nature of quantum computers, methods to sample from the stationary distribution have been of significant interest.
If a quantum algorithm can prepare the stationary distribution, sampling from this distribution can be done direct by simply measuring the state. 
The problem of preparing the stationary distribution of a Markov chain is a quantum state is called \textit{QSSamp}, as defined in~ \cite{Chakraborty_2020}. 

\subsection{Quantum algorithms for Markov chains}
Ergodic Markov chains $P$ have a mixing time of $\tilde{O}(1/\Delta)$, where $\Delta=|\lambda_1|-|\lambda_2|$ is the spectral gap of $P$ and $\lambda_1$ and $\lambda_2$ are the largest and second-largest eigenvalue of $P$. 
For stochastic matrices, the Frobenius-Perron theorem details that the largest eigenvalue of $P$ equals $\lambda_1=1$.
In addition, for random stochastic matrices of order $N$, the spectral gap scales as $O(\sqrt{N})$~\cite{Chafa__2010}, \cite{Horvat_2009}.

Richter conjectured the existence of a quantum algorithm for \textit{QSSamp} that could get $\epsilon$-close in time $\tilde{O}(1/\sqrt{\Delta})$, which would mean a quadratic speedup over the classical case~\cite{Richter_2007}. 
Further speedups have been theorised for a while, until the result by Aharonov and Ta-Shma in~\cite{aharonov2003}, which indicates that the existence of a more efficient quantum algorithm for \textit{QSSamp} is unlikely. 
Currently, it is believed that the conjectured $\tilde{O}(1/\sqrt{\Delta})$ is optimal.

Multiple algorithms attempted to attain this bound using various approaches , none of which proved successful to date.
Most of these approaches use the the quantum spatial search algorithm.
The quantum spatial search algorithm finds an element from a certain marked set, provided it starts from the stationary distribution. 
Classically, this requires a number of applications of the Markov chain known as the \textit{hitting time}, depending only on the Markov chain $P$ and the marked set.
Different quantum algorithms managed to devise quantum algorithms that obtain a quadratic improvement in the hitting time. 
Intuitively, running such a quantum spatial search algorithm in reverse would thus yield a preparation of the stationary distribution in this time. 
However, this reversal only achieves constant overlap with the stationary distribution. 
Quantum amplitude amplification or quantum phase estimation are required to end up with the stationary distribution. 
In 2019, Apers and Sarlette introduced fast-forwarding of a quantum Markov chain~\cite{Apers_Sarlette}.
This algorithm solves \textit{QSSamp} in a time scaling with the square root of the hitting time.
Although this is a nice result, it does not attain the bound conjectured by Richter, as the hitting time is at least as large as the mixing time of an ergodic, reversible Markov chain.

All of the above attempts approach \textit{QSSamp} with digital, or equivalently, gate-based quantum computing, which works by applying generic quantum gates to qubits
Recently, Chakraborth et al. proposed the first analog quantum algorithm to solve \textit{QSSamp}~\cite{Chakraborty_2020}. 
Analog quantum computing works by evolving specific initial states of qubits under suitable Hamiltonians for suitable evolution times. 
Their algorithm claims to find the stationary distribution in a time scaling with the sum of the square root of the hitting time and the square root of the classical mixing time.
As the hitting time is at least as large as the mixing time, this scales with the square root of the hitting time.
Although this also does not attain the conjectured bound by Richter, it does match the state-of-the-art obtained by discrete quantum algorithms.

In this section, an overview of the algorithm is given with the relevant details for the fine print in the next section.
The algorithm is presented as described in the original work, and no additional claims on correctness of the algorithm are made. 
In addition, some details are left out for clarity. 
A full description of the algorithm can be found in the original work~\cite{Chakraborty_2020}. 

The main building block of the algorithm is given by a Hamiltonian $H$ that encodes the Markov chain $P$.
The aspects of the Hamiltonian that are relevant for this work are that it works on the space of pairs of probability states; that the state $\ket{\pi,0}$ is an eigenvector with eigenvalue 0; and, that its other eigenvalues lie in the half-open interval~$(0,1]$.
At the start of the algorithm, the Hamiltonian is coupled to a pointer register, on which a momentum operator $\hat{p}$ can operate. 
The total Hamiltonian of the system is hence given by $\tilde{H}=H\otimes \hat{p}$.

It is well-known that, intuitively, the momentum operator can be seen as a generator of translation.
The pointer system allows to pick up a certain eigenvector when used in combination with post-selection. 
To be more precise, suppose that $H$ has eigenvalues $\lambda_{n^2} = 0<\lambda_{n^2-1}<\ldots<\lambda_1\leq 1$ with corresponding eigenvectors $\ket{u_{n^2}}, \ket{u_{n^2-1}}, \ldots, \ket{u_1}$. 
Any quantum state $\ket{\psi}$ can then be written in the eigendecomposition of $H$: $\ket{\psi} = \sum_{j=1}^{n^2}\alpha_j\ket{u_j}$.
Then, if the pointer register is initialised in the $\ket{x=0}$ state and $\tilde{H}$ is applied for time $t$, the following state is obtained:
\begin{equation*}
e^{-i\tilde{H}t}\ket{\psi}\otimes\ket{x=0} = \sum_{j=1}^{n^2}\alpha_j\ket{u_j}\ket{x=\lambda_j t}
\end{equation*}
Post-selecting on the state $\ket{x=0}$ then indeed yields the state $\ket{u_{n^2}}$ with eigenvalue $0$, which is equal to $\ket{\pi,0}$ for the chosen Hamiltonian.
Note that a positive value of $\alpha_{n^2}$ is required for the post-selection to succeed.

Lemma 4 and corollary 5 of Ref.~\cite{Chakraborty_2020} formalise this idea by proving that when $\tilde{H}$ is applied to the above state with a sufficiently large pointer register for a suitable time, the measured state is $\epsilon$-close to the eigenstate with eigenvalue 0. 
The time depends inverse quadratically on $|\alpha_{n^2}|$, which equals the inner product between the original state and the eigenstate with eigenvalue 0.
Hence, to efficiently prepare this 0-eigenstate, the initial state needs to have a sufficiently large, ideally constant, overlap with the state $\ket{\pi,0}$.

Unfortunately, it is not generally easy to prepare such a quantum state. 
The original paper solves this issue by applying corollary 5 twice.
First, corollary 5 is applied for a suitable interpolated Markov chain $P(s)$, to efficiently obtain a state $\epsilon$-close to a state with constant overlap with $\ket{\pi,0}$.
Given two Markov chains $P$ and $P'$, the interpolated Markov chain between them is given by $P(s)=(1-s)P+sP'$, controlled by the real parameter $s\in [0,1]$.
Afterwards, corollary 5 is applied with the original Markov chain, to efficiently obtain a state $\epsilon$-close to $\ket{\pi,0}$, as required.
For this idea to work, the above imposes two constraints on the 0-eigenstate of the Hamiltonian $\tilde{H}(s)$ corresponding to $P(s)$. 
First, the $0$-eigenstate of $\tilde{H}(s)$ should have constant overlap with $\ket{\pi,0}$ for the second application of corollary 5. 
Second, this 0-eigenstate should have constant overlap with the to-be-chosen initial state of the protocol for the first application of corollary 5. 
Specifically, this 0-eigenstate should have a constant overlap with an initial state that is easy to prepare.
Chakraborth et al. show that choosing an index $j$ such that the value $s^*=1-\pi_j/(1-\pi_j)$ lies in the open interval $(0,1)$ satisfies both constraints.
The corresponding initial state of the protocol then has to be chosen as $\ket{j,0}$, which can easily be prepared as it is a basis state of the quantum system.
The corresponding interpolated Markov chain is obtained by interpolating between the original Markov chain $P$ and the Markov chain $P'$ obtained by replacing all outgoing edges at state $j$ with self-loops, with interpolation parameter $s=s^*$.

In this way, the original work claims to solve the \textit{QSSamp} problem in time $\Theta(\frac{1}{\sqrt{\Delta(s^*)}}+\frac{1}{\sqrt{\Delta}}$), where $\Delta(s^*)$ is the spectral gap for the interpolated Markov chain $P(s^*)$.
Because of the choice of $s^*$, this can be shown to be equal to the sum of the square root of the classical mixing time and the square root of the classical hitting time.

\section{The Fine Print of the Algorithm}
\label{sec:fine_print}
In this section, the practical implementation of the algorithm by Chakraborty et al. is discussed, and it is shown that this speedup does not actually hold.

The original work proposed the algorithm as a theoretical tool to prepare the stationary state of a Markov chain with a theoretical speedup over its classical counterparts.
To run the algorithm as devised and hence obtain this speedup, the Markov chains $P$ and $P(s^*)$ need to be available as encoded by specific Hamiltonians $H$ and $H(s^*)$ as defined in the original work. 
Although this is not realistic for current near-term noisy (NISQ) hardware, it is possible that future fault-tolerant quantum (FTQ) devices will have these capabilities.

Assuming that FTQ devices indeed have these capabilities, we have to obtain a few parameters. 
To be precise, the algorithm requires the parameters $\varepsilon, s^*, \Delta$ and $\Delta(s^*)$, where the parameter $\varepsilon$ can be chosen freely by the user.
We will show that the complexity of determining these values exactly is larger than the complexity of the algorithm itself. 
In addition, we will show what the impact is of using approximations of these parameters.

\subsection{Parameter $s^*$}
To determine the parameter $s^*$, we first have to find an index $j$ such that $\pi_j$ lies within $(0,\frac12)$.
Then, the parameter $s^*$ is given by $s^*=1-\pi_j/(1-\pi_j)$. 
To prepare $s^*$, we thus have to know $\pi_j$ exactly, which corresponds to the stationary probability distribution we is precisely the distribution that is being prepared. 
Naturally, this results in circular logic, and can only be circumvented by finding an entry of the stationary distribution in a different way.
However, this different way will have a complexity larger than the algorithm by Chakraborty et al., as their algorithm tries to obtain a speedup over such alternative algorithms.

Note that there exist works that explain how to obtain good results even when only approximations for $s^*$ can be given~\cite{Krovi_2015}. 
However, in the case of this algorithm, even the slightest deviation in $s^*$ can completely diminish the promised speedup.
This can be shown by following the complexity derivation from Chakraborty et al. and replacing $s^*$ with an arbitrary value close to $s^*$.

Suppose that the original algorithm is run from an initial state $\ket{j,0}$ for which the corresponding variable $s^*$ is not exactly known. 
Instead, the algorithm is run with a value $s'$ replacing $s^*$.
The complexity of the full protocol is given by the sum of the complexities of the two parts of the protocol.
The complexity of the first part can be computed using Corollary 5, yielding a complexity of 
\[
\sqrt{\frac{1}{\Delta(s')|\alpha|^2}}\log(\frac{2}{\varepsilon|\alpha|^2})
\]
where $\alpha = \braket{j,0|v_n(s'),0}=\braket{j|v_n(s')}$.
Similarly, the complexity of the second part can be computed using Corollary 5, yielding a complexity of
\[
\sqrt{\frac{1}{\Delta|\beta|^2}}\log(\frac{2}{\varepsilon|\beta|^2})
\]
where $\beta = \braket{v_n(s'),0|\pi,0}=\braket{v_n(s')|\pi}$.
Using the expression for $\ket{v_n(s')}$ from Equation 13 in the original work, it follows that that
\[
\alpha = \sqrt{\frac{\pi_j}{1-s'(1-\pi_j)}} \quad \text{and} \quad \beta =\pi_j\sqrt{\frac{1}{1-s'(1-\pi_j)}}+\sum_{x\neq j}\pi_x\sqrt{\frac{1-s'}{1-s'(1-\pi_j)}}.
\]
Combined, the total complexity of the algorithm then equals
\[
\frac{1}{\sqrt{\Delta(s')}|\alpha|^2}\log(\frac{2}{\varepsilon|\alpha|^2})+\frac{1}{\sqrt{\Delta}|\beta|^2}\log(\frac{2}{\varepsilon|\beta|^2})
\]

\begin{figure}
    \centering
    \includegraphics[width=\linewidth]{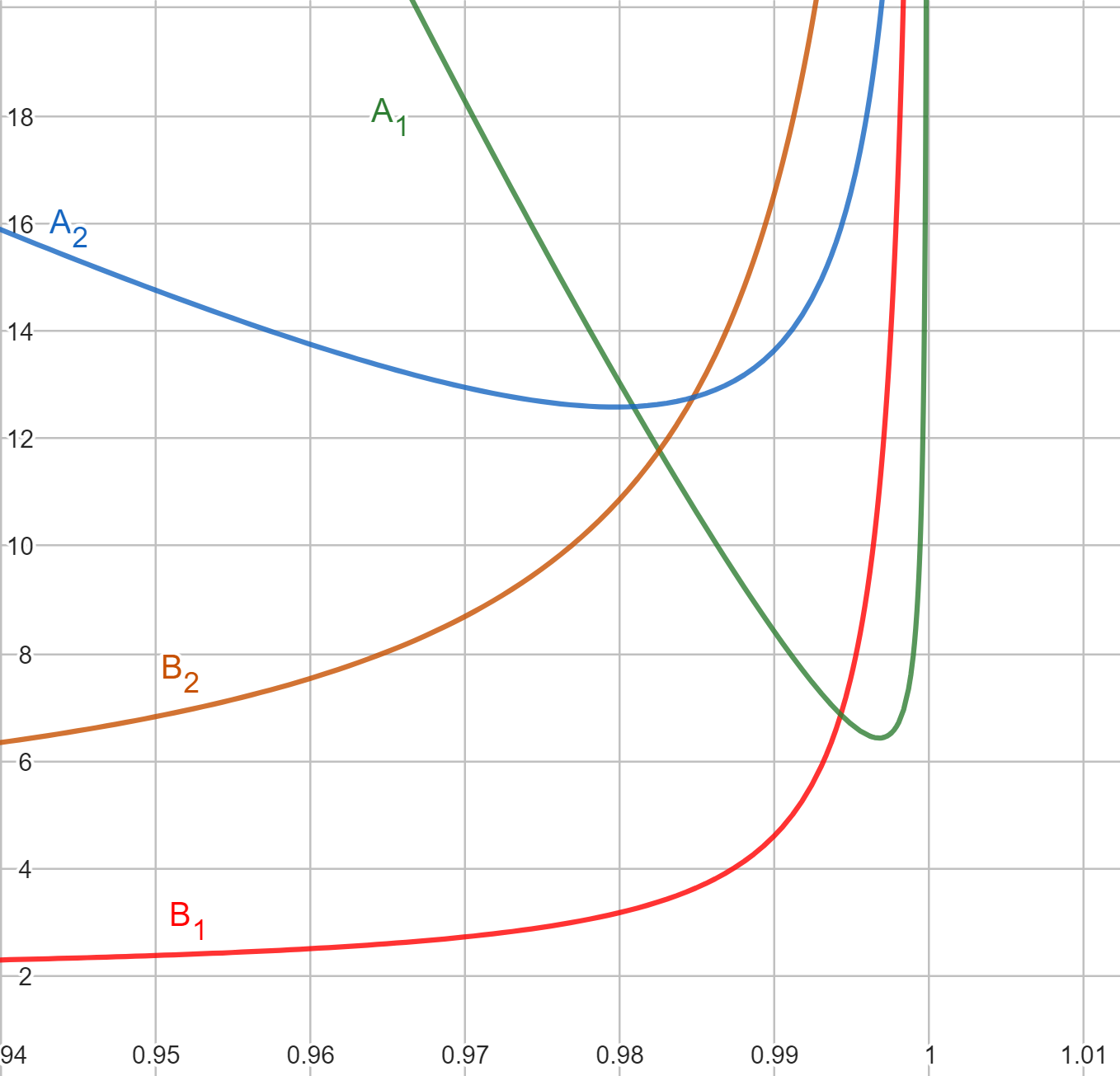}
    \caption{The values of $A$ and $B$ for two pairs of $(\epsilon, \pi_j)$, namely $(0.01, 0.1)$ and $(0.05, 0.5)$. The horizontal axis shows the value of $s'$, with the value of $s^*$ being at where the graph of $A$ is minimal (which is different for the two different cases).}
    \label{fig:figure}
\end{figure}

Note that, similar to the original algorithm, $\frac1\Delta$ scales as the mixing time.
In the original algorithm, $\frac1{\Delta(s^*)}$ scales as the hitting time. However, for approximate versions $s^*$, it holds that
\[
\frac{1}{\Delta(s')}\geq T_{hit}\frac{4}{1-|\alpha|^2}.
\]
This expression results from following the exact same steps in the original work that show that $\frac1{\Delta(s^*)}$ scales as the hitting time. All in all, the complexity of the full algorithm scales as
\[
A \sqrt{T_{hit}}+B\sqrt{T_{mix}}
\]
where
\[
A = \frac{1}{|\alpha|^2\sqrt{1-|\alpha|^2}}\log\left({\frac{2}{\varepsilon |\alpha|^2}}\right)
\quad \text{and} \quad
B = \frac{1}{|\beta|^2}\log\left({\frac{2}{\varepsilon|\beta|^2}}\right).
\]
Note that we omitted the factor $4$ in the $A$ term, as we only care for the asymptotic scaling. 
Recall that, if $s'=s^*$, the values $A$ and $B$ can be bounded from above by constants (in terms of the size of the matrix.
In the current situation however, this is not the case and the values depend on $\pi_j$.
To see how the value of $s'$ impacts the values of $A$ and $B$ in general, Figure~\ref{fig:figure} shows a diagram plotting the values of $A$ and $B$ for varying values of $s'$ and $\pi_j$.
The minimal value for $A$ corresponds to the situation where $s'=s^*$. 
Hence, even small deviations from the correct $s^*$ results in significant changes in the values of $A$ and $B$. 

\subsection{Parameters $\Delta$ and $\Delta(s^*)$}
Second, the parameters $\Delta$ and $\Delta(s^*)$ are considered, whose impact can also be significant.
These parameters are obtained in similar fashion, assuming knowledge of $s^*$ and access to the Markov chains $P$ and $P(s^*)$.
Recall that these parameters are defined as the spectral gap of the corresponding Markov chains.
Again, these are quantities are not directly available to the user when applying the algorithm.
From Ref.~\cite{hsu2017mixing}, it follows that the complexity to compute the spectral gap of a Markov chain is at least $O(\frac1{\Delta})$.
This complexity corresponds exactly the complexity of the classical mixing time, and that classical mixing time complexity is exactly the complexity the algorithm by Chakraborty et al. tries to improve upon.
Hence, the claimed speedup does not actually hold when these spectral gaps are not available beforehand, and in such cases the algorithm even achieves the same complexity as the original classical algorithm (up to logarithmic factors).

Using approximate values for the spectral gaps instead of the correct ones, affects the algorithm in various ways, depending on whether a value larger or smaller than the correct value is used, and also on which incorrect spectral gap is used. 

First, if a smaller value is used, it can be seen by following the same calculation as the original algorithm, that the correct state is obtained with at least the same precision. Indeed, a higher value for one of the spectral graphs just means that more pointer qubits are used, resulting in a higher overlap between the required states. However, this does result in an increased multiplicative complexity of the factor $O(\frac1{\sqrt{\Delta}})$.

Second, using a value higher than the correct spectral gap directly affects the performance of the algorithm.
The only impact an incorrect spectral gap has, is that in Lemma~4 of the original work, the overlap between the target and resulting state is smaller. The amount of overlap decreases linearly with the spectral gap. If a value twice as big is chosen as the correct spectral gap, no guarantee can be given on these two states having any overlap at all, and the algorithm is completely redundant. 

For smaller incorrect larger values, the impact on the algorithm differs depending on which spectral gap uses a higher value. 
If an incorrect value for $\Delta$ is used, the result is just that the overlap between the resulting state and target state of the full algorithm is smaller, if the original parameters are used. 
In other words, the performance of the algorithm decreases.
This can be fixed however: suppose the user uses a value of $\Delta'>\Delta$ instead of $\Delta$ such that $\Delta'/\Delta = C < 2$.
Then if one runs the second part of the algorithm with $\lceil{\log_{2/C}(4/\varepsilon)}\rceil$ copies of the pointer register instead of $\lceil{\log_{2}(4/\varepsilon)}\rceil$ copies, it still obtains the result with the same precision.
However, the complexity of the second part of the algorithm, i.e., the $\sqrt{\frac{1}{\Delta}}$ term, increases with a factor of roughly $\log_{2/C}(2)$. If the value of $C$ can be bounded at least a constant away from $2$, the same complexity holds, albeit with a (considerably) worse constant. 
Thus, if $\Delta$ can be approximated within a constant factor smaller than 2 from above, the algorithm runs with the same precision, but a higher complexity.

If an incorrect value for $\Delta(s^*)$ is used, the impact is slightly different. Again, suppose a value of $\Delta'>\Delta(s^*)$ is used, with $\Delta'/\Delta(s^*)=C<2$. 
Just like in the previous paragraph, the same precision of the algorithm can be assured by increasing the complexity. 
This would ensure that the resulting state after step 1 is within the distance to the target state required after step 1. 
Just like before, this requires an increase in the $\sqrt{\frac{1}{\Delta(s^*)}}$ term in the complexity with a factor of roughly $\log_{2/C}(2)$. 
There is also another option here, namely running this first part of the algorithm with the original parameters. The state obtained after step 1 of the algorithm then has a smaller overlap with the target state after step 1 of the algorithm than intended, namely an overlap of $\delta = (\varepsilon/4)^{\log_{1/2}(C/2)}$. 
If step 2 of the algorithm would then be run from this state, this would require a complexity of $\frac{1}{\Delta\delta}\log(\frac{1}{\delta})$, assuming a correct value for $\Delta$ (or for approximate versions of $\Delta$ with the same complexity increase as before).
Again, if the value of $C$ can be bounded at least a constant away from $2$, the same complexity holds, albeit with a (considerably) worse constant.
Depending on the exact value of $C$ and $\varepsilon$, the first or second alternative gives more beneficial scaling factors.

In conclusion, if the spectral gaps can be approximated within a constant factor smaller than 2, the algorithm still performs with the same complexity, albeit with (considerably) worse constants. 
However, to the best of our knowledge, the best algorithm to approximate the spectral gap within a factor of 2 is to simply compute the spectral gap exactly. 
It was mentioned before that other approximations (to higher factors) for the spectral gaps nullify the speedup claimed by the original algorithm. 
In summary, there are no algorithms that can sufficiently approximate the spectral gaps such that the speedup of the algorithm remains intact, unless the spectral gap is computed manually. 
As computing the spectral gap manually requires a complexity similar to the classical mixing time, this completely negates the speedup presented by the algorithm, and instead the algorithm will have the same complexity as computing the classical mixing time.

\subsection{Instances in which speedup is preserved}
Altogether, it can be concluded that in the general case, the algorithm by Chakraborty et al. receives at best the same complexity as the classical mixing algorithm.
Of course, it should be noted that there exist instances in which the required parameters are naturally available.
It is however expected that there are only a limited amount of such situations, as both $s^*$ as well as the spectral gaps of two different Markov chains have to be naturally available to the user.
In particular, to the best of our knowledge, no natural examples exist where these values are all known to the user.

\section{Discussion}
\label{sec:discussion}
In this work, we discussed the fine print of the first analog quantum algorithm to prepare the stationary distribution of a Markov chain.
The original work claims a speedup over the classical state-of-the-art, and even claims to match the (digital) quantums state-of-the-art.
This work, in constrast, shows that this speedup actually depends on the knowledge of a couple of specific variables.
Specifically, it is shown that to prepare these variables, a complexity is required which is similar to the complexity of classically preparing the stationary distribution.
From this, the complexity of the practical implementation of the algorithm was computed.
From this complexity, it can be concluded the promised speedup offered by this algorithm does not hold in practice.

It should be noted that the algorithm in Ref.~\cite{Chakraborty_2020} still proves useful in Markov chain problems, yet, with a worse complexity than originally claimed. 
In addition, it might be possible to use their ideas in other algorithms that do obtain this claimed speedup.

With this fine print, this work adds to the line of research of considering the practical applicability of quantum algorithms, started by the seminal work of Aaronson~\cite{Aaronson2015}, later extend by other works, such as~\cite{Neumann:2019}.
Just like these previous works, this work shows the importance of this line of research, as practical implementation is often very different from theoretical description.
For future work, it would be interesting to consider the fine print of other algorithms and hence continue this important line of research.

The goal of the quantum computing field should be to bring society further. 
To achieve this, more and more quantum algorithms are being proposed to obtain speedups over state-the-art classical solvers, or to solve classically intractable problems.
Most of these algorithms offer novel insights and indeed indicate speedups over classical best practices.
However, by omitting the so-called fine print of these algorithms, there cannot be sufficient trust in that these theoretical speedups actually transfer to practical implementation.
Quantum computing can only bring society further by being explicit about the full applicability and fine print of quantum algorithms. 

\bibliographystyle{IEEEtran}
\bibliography{references}
\end{document}